\documentclass[aps,prl,reprint]{revtex4-1}
\usepackage{graphicx}

\begin{document}
\title{Stochastic synchronization of genetic oscillators induced by miRNA}
\author{Saurabh Kumar Sharma$^1$}
\author{Shakti Nath Singh$^1$}
\author{Md. Jahoor Alam$^1$}
\author{R.K. Brojen Singh$^{1}$}
\email{brojen@mail.jnu.ac.in}
\affiliation{$^1$Centre for Complex Systems Studies, School of Computational and Integrative Sciences, Jawaharlal Nehru University, New Delhi-110067, India.}

\begin{abstract}
We study two important roles of $miRNA$ as stress inducer and synchronizing agent in regulating diffusively coupled genetic oscillators within stochastic formalism and competition between them. We identify optimal value of coupling constant at which the rate of synchronization is strongest, below this value sychronizing activity dominates stress activity, and above it stress activity destroys synchronizing pattern. The concentration of $miRNA$ maintained in individual oscillator modulates the synchronization activity achieved by diffusing $miRNA$.
\end{abstract}


\maketitle

\subsection{Introduction}

$miRNA$s are class of non-coding small and single stranded RNA molecules of length about 22-24 nucleotide bases\cite{bus,amb}, and has variety of significant role in development and various cellular functions \cite{amb1,bus}. $miRNA$s are transcried from non-coding genomic regions and is generated through a multi-step process starting from the nucleus and ends in the cytoplasm \cite{car}. $miRNA$s has significant role in regulating gene and protein expression \cite{peg} and regulates various biological processes, such as immune response, metabolism, cell-cycle control, viral replication, stem cell differentiation and human development \cite{eth}. It is also reported that most of the $miRNA$s are conserved across multiple species, which made it evolutionary important molecules as modulators of critical biological pathways \cite{chen}. Further, the $miRNA$ expression or function was found to be significantly altered in many disease states, such as cancer, metabolic diseases, inflammatory diseases, fibrosis etc \cite{klo}. 

Genetic oscillator, which can able to generate 24 hours cycle, is a model oscillator of circadian clock \cite{jos} constructed on the principle of self-sustained transcriptional and translational feedback loops \cite{dun,gal,zhe}. The circadian clock involves various clock genes and proteins which enable the organisms to anticipate environmental changes or segregate in time-incompatible processes \cite{til}. The clock mechanism consists of positive regulating components (regulators of clock genes) and negative regulating components which interact physically with each other to produce heterodimers of them and later shuttled back to the nucleus to repress the positive components \cite{dar,kum,lee1,ala}. Genome wide expression analysis from several micro-array studies in different organisms showed that the expression of as much as (5-10)\% of transcripts in a specific tissue oscillates in a circadian manner \cite{mcd,akh,ued,kee}.

Recently, studies on micro-array data of Drosophila heads indicated that the expression of 78 $miRNA$s from flies entrained to light-dark cycles \cite{yan}. The expression of these $miRNA$s were validated by qPCR experiments, which also revealed participation of these $miRNA$s in circadian regulation \cite{yang}. Further, it was also experimentally proved the similar roles of $miRNA$ in higher organisms (rat) showing variation in $miRNA$ level after sleep deprivation in the rat’s brain triggered modest magnitude of fold-change \cite{dav}. It is also believed that there could be a threshold level of $miRNA$ after which it start regulating circadian rhythm and this mechanism may respond to various factors, for example, temperature, environmental fluctuations etc \cite{yan}. However, the way how $miRNA$ regulate circadian rhythm and mechanism of inducing stress to the cellular system is not fully studied yet.

Recently, $miRNA$s have been detected in extracellular body fluids with significantly high concentration and found to be more stable outside the cell \cite{kai}. This experimental and micro-array analysis report suggest the possibility of circulating $miRNA$ through the body fluid which could serve as useful clinical biomarker \cite{eth}. More stability of $miRNA$ in extracellular medium and its circulating ability in the body fluid suggests its important role as signaling molecule in cell to cell communication \cite{kai}, however, it is still an open question.

The biochemical network system can be studied using stochastic approach based on Master equation formalism which involves decay and creation of molecular species due to random molecular interaction \cite{gil,kam}. The information processing among such coupled identical stochastic systems can be studied using the concept of stochastic synchronization \cite{pik,nan}. Such study of stochastic synchronization can probably highlight some important insights of complicated signal processing and cell to cell communication at fundamental level \cite{ram}. In this work we try to study some of the basic questions of $miRNA$ and its important roles in regulating genetic oscillator.

\subsection{Model of miRNA coupled genetic networks}

The genetic oscillator model we consider (Fig. 1) is the extension of the Vilar's model \cite{jos} which incorporates the impact of $miRNA$ \cite{ami}. The model consists of two genes, activator (with promoters $D_A$ and $D_A^\prime$) and repressor (with promoters $D_R$ and $D_R^\prime$) which are transcribed and translated into their respective mRNAs ($M_A$ and $M_R$) and proteins ($A$ and $R$) respectively \cite{jos}. The interaction of proteins $A$ and $R$ leads to the formation of the complex $C$ and $C$ degrades to $R$ with different rate constants. The list of molecular species and the set of reactions involved in the biochemical network of genetic oscillator with the kinetic laws of individual reaction with the respecive rate constants are listed in Table 1 and Table 2 respectively.

The introduction of $miRNA$ in the regulation of genetic oscillator can be done by introducing four basic reactions in the network \cite{ami}. The $miRNA$ allows to interact with repressor mRNA ($M_R$) to form a complex $C_{RISC}$ with a rate constant $k_{21}$ and $C_{RISC}$ degrades with another rate constant $k_{22}$ (we did not consider the reversible reaction of $C_{RISC}$). The synthesis and degradation of $miRNA$ in the system take place with different rate constants (Table 2).
\begin{table}
\begin{center}
{\bf Table 1 - List of molecular species} 
\begin{tabular}{|l|p{2cm}|p{3cm}|p{2cm}|}
 \hline \multicolumn{4}{}{} \\ \hline

\bf{ S.No.}    & \bf{Molecular Species} & \bf{Description}                          &  \bf{Notation} \\ \hline
1.             &    $D_A$               & Activator gene without activator protein &  $Y_1$         \\ \hline
2.	       &    $D'_A$              & Activator gene with activator protein    &  $Y_2$         \\ \hline
3.             &    $A$                 & Activator protein        		    &  $Y_3$         \\ \hline
4.	       &    $D_R$               & Repressor gene without activator protein  &  $Y_4$         \\ \hline
5.             &    $D'_R$              & Repressor genes with activator protein    &  $Y_5$         \\ \hline
6.	       &    $M_R$               & mRNA of the repressor protein             &  $Y_6$         \\ \hline
7.             &    $M_A$               & mRNA of the activator protein             &  $Y_7$         \\ \hline
8.	       &    $R$                 & Repressor protein	         	    &  $Y_8$         \\ \hline
9.             &    $C$                 & Activator Repressor complex               &  $Y_9$         \\ \hline
10.	       &    $m$                 & microRNA			            &  $Y_{10}$      \\ \hline	
11.            &    $C_{RISC}$          & Complex of microRNA with mRNA	            &  $Y_{11}$      \\ \hline
\end{tabular}
\end{center}
\end{table}


\begin{table*}
\begin{center}
{\bf Table 2 List of chemical reaction, propensity function and their rate constant} 
\begin{tabular}{|l|p{2.5cm}|p{4.5cm}|p{2.0cm}|p{1cm}|}
\hline \multicolumn{5}{}{}\\ \hline

1  & $Y_2\stackrel{k_{1}}{\longrightarrow}Y_1$       	  & $a_1$=$k_1Y_2$                  & $k_1$=50 &       \cite{ami,jos}\\ \hline
2  & $Y_1+Y_3\stackrel{k_{2}}{\longrightarrow}Y_2$   	  & $a_2$=$k_2(Y_1Y_3$)          & $k_2$=1 &        \cite{ami,jos}\\ \hline
3  & $Y_5\stackrel{k_{3}}{\longrightarrow}Y_4$       	  & $a_3$=$k_3Y_5$                  & $k_3$=100 &      \cite{ami,jos}\\ \hline
4  & $Y_4+Y_3\stackrel{k_{4}}{\longrightarrow}Y_5$   	  & $a_4$=$k_4(Y_4Y_3$)          & $k_4$=1 &        \cite{ami,jos}\\ \hline
5  & $Y_5\stackrel{k_{5}}{\longrightarrow}Y_6+Y_5$        & $a_5$=$k_5Y_5$                  & $k_5$=50 &       \cite{ami,jos}\\ \hline
6  & $Y_4\stackrel{k_{6}}{\longrightarrow}Y_6+Y_4$        & $a_6$=$k_6Y_4$                  & $k_6$=0.01 &     \cite{ami,jos}\\ \hline
7  & $Y_6\stackrel{k_{7}}{\longrightarrow}\phi$           & $a_7$=$k_7Y_6$                  & $k_7$=0.5 &      \cite{ami,jos}\\ \hline
8  & $Y_2\stackrel{k_{8}}{\longrightarrow}Y_7+Y_2$        & $a_8$=$k_8Y_2$                  & $k_8$=500 &      \cite{ami,jos}\\ \hline
9  & $Y_1\stackrel{k_{9}}{\longrightarrow}Y_7+Y_1$        & $a_9$=$k_9Y_1$                  & $k_9$=50 &       \cite{ami,jos}\\ \hline
10 & $Y_7\stackrel{k_{10}}{\longrightarrow}\phi$          & $a_{10}$=$k_{10}Y_7$            & $k_{10}$=10 &    \cite{ami,jos}\\ \hline
11 & $Y_6\stackrel{k_{11}}{\longrightarrow}Y_6+Y_8$       & $a_{11}$=$k_{11}Y_6$            & $k_{11}$=5 &     \cite{ami,jos}\\ \hline
12 & $Y_8\stackrel{k_{12}}{\longrightarrow}\phi$          & $a_{12}$=$k_{12}Y_8$            & $k_{12}$=0.2 &   \cite{ami,jos}\\ \hline
13 & $Y_9\stackrel{k_{13}}{\longrightarrow}Y_8$           & $a_{13}$=$k_{13}Y_9$            & $k_{13}$=1 &     \cite{ami,jos}\\ \hline
14 & $Y_7\stackrel{k_{14}}{\longrightarrow}Y_7+Y_3$       & $a_{14}$=$k_{14}Y_7$            & $k_{14}$=50 &    \cite{ami,jos}\\ \hline
15 & $Y_2\stackrel{k_{15}}{\longrightarrow}Y_2+Y_3$       & $a_{15}$=$k_{15}Y_2$            & $k_{15}$=50 &    \cite{ami,jos}\\ \hline
16 & $Y_5\stackrel{k_{16}}{\longrightarrow}Y_5+Y_3$       & $a_{16}$=$k_{16}Y_5$            & $k_{16}$=100 &   \cite{ami,jos}\\ \hline
17 & $Y_3\stackrel{k_{17}}{\longrightarrow}\phi$          & $a_{17}$=$k_{17}Y_3$            & $k_{17}$=1 &     \cite{ami,jos}\\ \hline
18 & $Y_8+Y_3\stackrel{k_{18}}{\longrightarrow}Y_9$       & $a_{18}$=$k_{18}(Y_8Y_3$)    & $k_{18}$=2 &     \cite{ami,jos}\\ \hline
19 & $\phi\stackrel{k_{19}}{\longrightarrow}Y_{10}$       & $a_{19}$=$k_{19}$                & $k_{19}$=20 &    \cite{ami}\\ \hline
20 & $Y_{10}\stackrel{k_{20}}{\longrightarrow}\phi$       & $a_{20}$=$k_{20}Y_{10}$         & $k_{20}$=0.029 & \cite{ami}\\ \hline
21 & $Y_{10}+Y_6\stackrel{k_{21}}{\longrightarrow}Y_{11}$ & $a_{21}$=$k_{21}(Y_{10}Y_6$) & $k_{21}$=6 &     \cite{ami}\\ \hline
22 & $Y_{11}\stackrel{k_{22}}{\longrightarrow}\phi$       & $a_{22}$=$k_{22}Y_{11}$         & $k_{22}$=0.6 &   \cite{ami}\\ \hline
\end{tabular}
\end{center}
\end{table*}

\begin{figure}
\begin{center}
\includegraphics[height=330 pt,width=270 pt]{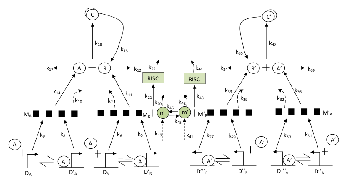}
\caption{Schematic diagram showing the coupling of identical genetic oscillators.}
\end{center}
\end{figure}

Since $miRNA$ is quite stable in extracellular medium, it can be taken as important signaling molecule which can carry information from one cell to another \cite{kai}. We then take two identical oscillators diffusively coupled via $miRNA$. This modeling can be done by constructing a larger system ($S$) where the two genetic oscillators are sub-systems ($S_1$ and $S_2$) and introducing two extra reactions corresponding to the in and out diffusion of $miRNA$ between the two sub-systems, given by,
\begin{eqnarray}
miRNA\stackrel{\epsilon}\rightarrow miRNA^\prime\\
miRNA^\prime\stackrel{\epsilon^\prime}\rightarrow miRNA
\end{eqnarray}
where, $\epsilon$ and $\epsilon^\prime$ are coupling rate constants corresponding to the two reations. The larger system ($S=S_1\cup S_2$) consists of forty six reactions and twenty two molecular species.

We use stochastic simulation algorithm (SSA) due to Gillespie \cite{gil} to simulate the set of reactions using the parameter values listed in the Table 2. The SSA is a Monte carlo type of simulation algorithm based on two fundamental questions, at what time which reaction will get fired, and accordingly the population of each molecular species will be updated as a function of time \cite{gil}. Thus one can trace the time evolution of the population states of the system in population-time space.
\begin{figure}
\begin{center}
\includegraphics[height=200 pt, width=200 pt]{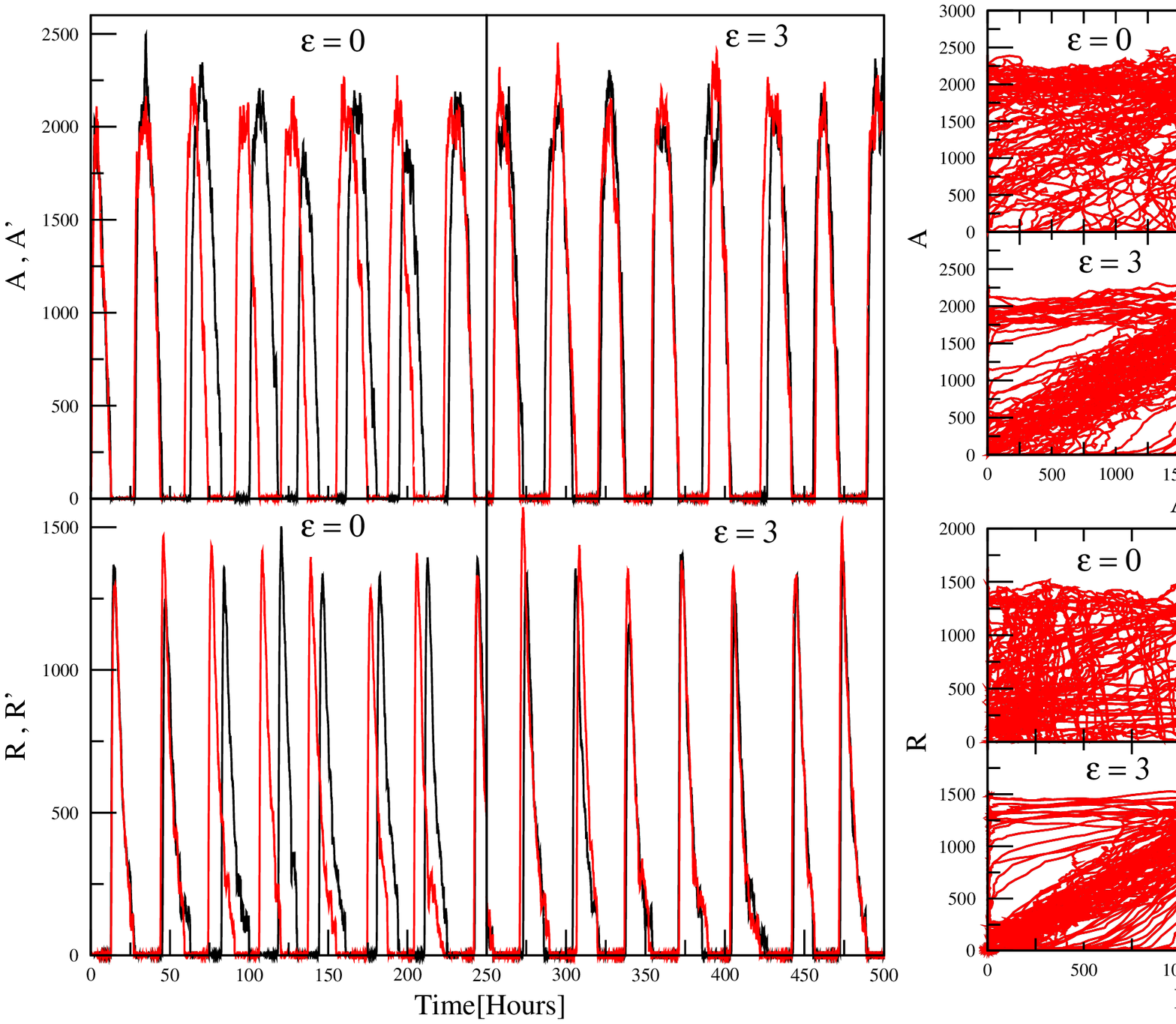}
\caption{Synchronization of two stochastic genetic oscillators: (a) Time series of ($A, A^\prime$) and ($R, R^\prime$) where coupling is switched on at 250 hours, (b) two dimensional recurrence plots of the coupled oscillators.}
\end{center}
\end{figure}

\subsection{Results and discussion}

The simulation of the coupled genetic oscillators via $miRNA$ is done using SSA by taking $\epsilon=\epsilon^\prime$ and reaction channels with corresponding parameter values listed in Table 2. The results are discussed as in the following.

$miRNA~induced~synchronization.-$
We first present the simulation results of synchronization induced by $miRNA$ (Fig. 2) where coupling is switched on at 250 hours. When $\epsilon=\epsilon^\prime=0$, the two oscillators are uncoupled and their behaviours are independent of each other in the time series of $A$ and $R$. The random distribution of points in the two dimensional recurrence plots of ($A$ and $A^\prime$) and ($R$ and $R^\prime$) show the uncorrelated behaviour of the two oscillators (Fig. 2 first and third panels). However, if the value of $\epsilon$ increases, the dynamics of the two oscillators start correlated to each other and become strongest when $\epsilon=3$ (Fig. 2). The points in two dimensional recurrence plots in $A$ (Fig. 2 second panel) and $R$ (Fig. 2 fourth panel) become concentrated along the diagonal showing strong correlation. The distribution of the points along the diagonal have a thickness due to stochastic noise in the system which have destructive nature in the signal processing.

$Random~coupling~induced~synchronization$
The random coupling between the two oscillators can be done by putting $\epsilon=\epsilon^\prime=Cr$, where, $r$ is the random number and $C$ is the magnitude of random fluctuation in random coupling. In this case of random coupling scheme, the oscillators become synchronized at $C=8$ (Fig. 3) which is verified by two dimensional recurrence plots for both $A$ and $R$ variables. This means that if the oscillators are coupled with random diffusion of $miRNA$, the synchronization is achieved at higher values of coupling strength. The disorderness induced by random diffusion of $miRNA$ from one oscillator to another probably may trigger change in the correlation between them which may need higher value of coupling strength to get them synchronized.  
\begin{figure}
\begin{center}
\includegraphics[height=220 pt, width=220 pt]{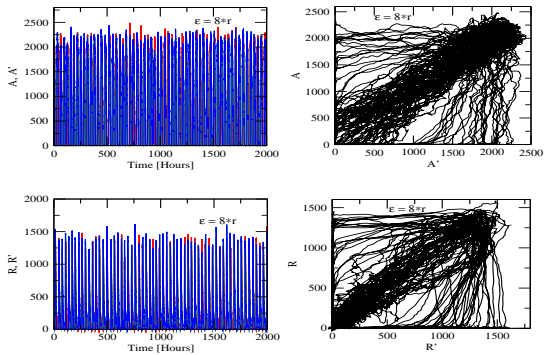}
\caption{Synchronization of two identical stochastic oscillators with random diffusive coupling mechanism.}
\end{center}
\end{figure}

$miRNA~as~synchronizer~and~stress~inducer.-$
The role of $miRNA$ in the genetic network could be different depending on the concentration of $miRNA$ diffused in the system (depending on the coupling constant $\epsilon$). If $\epsilon$ is small the two oscillators are not correlated and at the same time the stress induced by the diffused $miRNA$ to the individual oscillator is also small and therefore the genetic oscillators behave as near normal and uncoupled oscillators. Then as $\epsilon$ increases the oscillators start correlated each other and strongly synchronized at $\epsilon=3$ (Fig. 2, Fig. 4). However, if the value of $\epsilon$ is increased further then the two oscillators start uncorrelated to each other and desynchronized at $\epsilon=15$ (Fig. 4 lower left panels) and above, which is supported by two dimensional recurrence plots (Fig. 4 lower right panels) where points randomly scattered away from diagonal. This exhibited uncorrelated behaviour of the oscillators at higher values of $\epsilon$ could be due to stronger activity of stress over synchronizing activity. The results reveal that synchronizing activity controls correlation between coupled genetic oscillators for moderate values of $\epsilon$, however, stress activity takes over synchronizing activity at higher values of $\epsilon$ (excess diffused $miRNA$).
\begin{figure}
\begin{center}
\includegraphics[height=220 pt, width=220 pt]{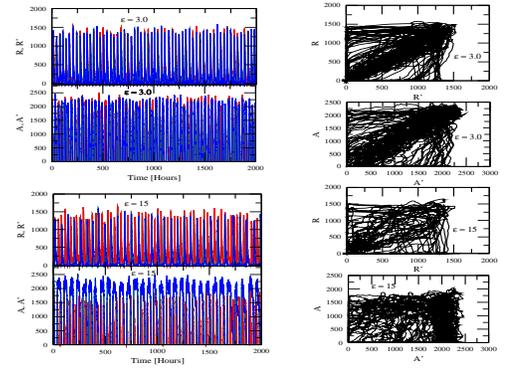}
\caption{Plots showing the competition between stress and synchronizing activities: (a) the synchronization become strongest for $\epsilon=3$, below it synchronization activity dominates stress activity, (b) above $\epsilon=3$ stress activity dominates synchronization activity and synchronization is destroyed at $\epsilon=15$.}
\end{center}
\end{figure}

$Rate~of~synchronization~induced~by-miRNA$
We now study the impact of concentration of $miRNA$ synthesized inside the system (which is proportional to $_{19}$) in each oscillator on the rate of synchronization. We took various values of $k_{19}$ (18, 19, 21 and 22) and calculated the values of $\epsilon$ where the synchronization takes place (Fig. 5 upper left panel). The error bars indicate the range of $\epsilon$. We can not take large values of $k_{19}$ (large concentration of $miRNA$) because excess stress induced by $miRNA$ will destroy the oscillatory behaviour of genetic oscillator \cite{ami}. The results show that $\epsilon$ fluctuates on varying $k_{19}$ but increases on an average.

The increase in $k_{20}$ (degradation rate of $miRNA$) first compel $\epsilon$ to increase and then slowly decreases as $k_{20}$ increases further (Fig. 5 right upper panel). The result indicates that the value of $k_{20}$ should have optimal value where $\epsilon$ attains maximum value. The impact of $k_{21}$ (rate constant of formation of complex $C_{RISC}$) on $\epsilon$ is to increase $\epsilon$ as $k_{21}$ increases (Fig. 5 lower left panel). On the other hand $\epsilon$ decreases as $k_{22}$ (degradation rate of $C_{RISC}$) increases (Fig. 5 lower right panel). The results indicate that the rate constants related to $miRNA$ interaction to genetic oscillator should have optimal values in order to maintain balance between stress and synchronizing activities.
\begin{figure}
\begin{center}
\includegraphics[height=200 pt]{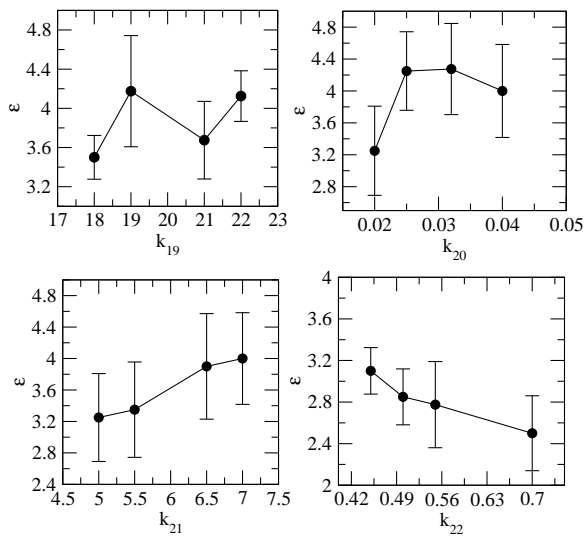}
\caption{Plots showing modulation of rate of synchronization ($\epsilon$) by $miRNA$ concentration available in each individual oscillator.}
\end{center}
\end{figure}

\subsection{Conclusion} 

In this study, $miRNA$ has been shown as an important signaling molecule which is important for cellular communication. We found two contrast but important roles of $miRNA$ which regulate coupled genetic oscillators. Depending on the availability of the diffused $miRNA$ between two coupled genetic oscillators via $miRNA$, the way of information processing between two genetic oscillators is different. The correlation between two coupled oscillators is found to be maximum (strongest synchronization) for optimal concentration of $miRNA$, above this value the correlation get destroyed. This condition is due to excess increased in $miRNA$ which induce excess stress to the system. Therefore there is always competition between synchronizing and stress inducing activities in the system based on the level of $miRNA$ concentration. This regulating mechanism of $miRNA$ could trigger various important roles of it in understanding complex information processing.

In coupled genetic oscillators, the available $miRNA$ concentration in each oscillator is due to $miRNA$ diffused and synthesized in the cell. The synthesized $miRNA$ in the cell could able to modulate genetic oscillator \cite{ami} as well as control the rate of synchronization or information processing. This important regulating mechanism of $miRNA$ could open up key ways to control various diseases and identify important disease regulators.

$Acknowledgments.-$ This work is financially supported by Department of Science and Technology (DST), New Delhi, India under sanction no. SB/S2/HEP-034/2012.

\end{document}